\documentclass{blois}

\def\be{\begin{equation}}
\def\ee{\end{equation}}
\def\bea{\begin{eqnarray}}
\def\eea{\end{eqnarray}}

\newcommand{\Photo}{}

\begin{document}
\vspace*{4cm}
\title{BLOIS 2019: HIGHLIGHTS  AND OUTLOOK}

\author{GRACIELA B. GELMINI }

\address{Department of Physics and Astronomy, University of California, Los Angeles,\\ CA 90095-1547, USA
}

\maketitle\abstracts{
This is an idiosyncratic account of the main results presented at the 31st Rencontres de Blois, which took place from June 2nd to June 7th, 2019 in the Castle of Blois, France. 
}

\section{The two Standard Models}

This conference is characterized by including all aspects, theoretical and experimental/observational of  particle physics and cosmology/astrophysics, embracing the true interdisciplinary nature of research in these fields.

Particle physics and cosmology/astrophysics are inextricably intertwined. Since the 1970's onwards  two standard models were developed, one for particle physics and one for cosmology.  The Standard Model (SM) of elementary particle physics, provides a complete description of all the known fundamental interactions except gravity (so it cannot be valid beyond the Planck energy scale).  The standard model of
cosmology is the Big Bang model, of an adiabatically expanding radiation dominated early 
Universe, in which the content of the Universe is given by the particles of the SM and whose subsequent evolution depends on the equations of state of the different components of the energy density.  

Both standard models
 combined provide a common framework to try to understand the still open problems in both disciplines, particle physics and cosmology, such as: the flatness and horizon problems of the Universe through inflation,
the formation of structure in the Universe (through primordial inhomogeneities generated as quantum fluctuations during inflation and growth which depends on the content of the Universe),  the matter-antimatter asymmetry of the Universe and
what the dark matter and the dark energy (which together comprise 95\% of the content of the Universe) consists of.  

We are in an era of major discoveries which confirm the two standard models  and we love anomalies to challenge these models and make progress. The conference highlighted the major recent discoveries and addressed the status of present anomalies in all subfields of particle physics and cosmology: tests of the SM, 
the characterization of the Higgs boson, theories of and searches for physics beyond the SM (BSM),
heavy flavour physics, neutrino physics including astrophysical neutrinos,
astroparticle physics, dark matter, dark energy and cosmology in general.

\section{Major Discoveries in Recent Years}

Three of the most significant discoveries of the last decade were befittingly highlighted by the three inaugural talks of the conference:  ``The Future of Cosmology \&HEP" by George Smoot,  the ``Status of research on  gravitational waves: from kilo-Hz to name-Hz" by Stanislav Babak and  ``Ten Years of LHC- Highlights, Challenges and Opportunities"  by Marumi Kado.  The three major advances these talks referred to were (in chronological order) the 2012 discovery at the LHC of the Higgs boson, the last piece which completed the SM,  the 2013 Planck data release (followed by other two in 2015 and 2018) which provided a determination with unprecedented precision of the characteristics of CMB anisotropies, and the  2015 detection of gravitational waves by LIGO and VIRGO, which gave us new eyes with which to observe the Universe. 

These three would for sure be in everybody's list of major advancements in particle physics and cosmology in the last ten years,  although others could be added (e.g. the first observation of PeV energy extragalactic neutrinos in IceCube in 2014 and the first multi-messenger observation with gravitational waves and photons of the merger of two neutron-stars in 2017).  Each one of the three advances were so transformative that determined a before and an after in our field.

\subsection{Before and after the Higgs boson discovery}

Before the Higgs boson discovery,  physicists knew that there should be something the LHC would find at the electroweak unification scale, i.e. at partonic energy scales of $O$(100 GeV), whether it was the Higgs boson or not. The reason is that without the Higgs boson the $W W \to W W$ scattering cross section grows monotonically with increasing energy, and at a point below the TeV scale produces probabilities greater than one, which points out that something is missing. The argument that the LHC had to be able to see something,  is called  the  ``No-lose Theorem" (although the argument was first formulated with a lepton collider in mind~\cite{Espinosa:1998xj}). Before the discovery of the Higgs boson as a narrow resonance, there were guaranteed discoveries at the electroweak unification scale. 

After the discovery, just the Higgs and nothing else reachable at the LHC is a possibility, so there is no guaranteed discovery. Alain Blondel in his talk at the conference put it succinctly as  ``we have no scale", meaning there is no known energy scale at which BSM physics must appear. 

 However,  the SM has open problems that encourage us to keep looking  for BSM physics.
One of them is the electroweak hierarchy problem which arises because elementary scalars, such as the Higgs, are quadratically sensitive to physics at higher scales (the Planck scale, 10$^{19}$ GeV,  if there are no other lower scales of BSM physics).  The  Higgs mass value and all other available data  leave open  the possibility of supersymmetry (SUSY) at a scale reachable at the LHC, although not in the form of ``light-vanilla" supersymmetric models.  They also allow for Higgs composite models, in which the Higgs is a pseudo-Nambu Goldstone boson or  a dilaton, although many composite models have been rejected (such as those  ``higgless" and those with many new particles at the electroweak scale),  and for other models too.

Another open problem of the SM  is the stability of the Higgs potential vacuum, because the quartic self coupling in the potential may go negative at a scale close to 10$^{11}$ GeV. This problem was addressed by Oleg Levedev at the conference. 

Another clear indication of the incompleteness of the SM is that the active neutrinos, which are massless within the SM, do have non-zero mass (at least two of them, since there are two  different flavor oscillations) as discussed by Mu-Chun Chen at the conference.

These reasons lead us to believe that  the SM may not be the correct theory up to the Planck energy scale, but BSM physics may start at much lower energy scales. This can be tested experimentally through the direct observation of new particles,  or  the observation of  new phenomena (e.g. new neutrino oscillations beyond the two we know, new sources of CP violation, proton decay...), or the observation of deviations from precise predictions of the SM.

Marumi Kado gave a comprehensive review of the many searches for BSM physics at the two runs of the LHC in  the last 10 years. Run 1 had center of mass energies of 7 and 8 TeV and reached an integrated luminosity of  approximately 20 fb$^{-1}$ (for ATLAS and CMS). In the Run 2 the energy was  13 TeV and the integrated luminosity was about 140 fb$^{-1}$ (for ATLAS and CMS). After several  upgrades to the machine and detectors, Run 3 will start at the end of 2020  and extend until the end of 2023,  at 14 TeV center of mass energy with the aim of reaching 300 fb$^{-1}$ of integrated luminosity, after which the LHC will shut down again for a major upgrade. The High Luminosity phase, HL-LHC,  will start in 2026 again at 14 TeV center of mass energy,  and will reach a luminosity of 3000 fb$^{-1}$, running possibly until 2038. 

Many of the results obtained so far at the LHC were reviewed at the conference, both in plenary and parallel session talks (such as those on Higgs couplings by Marco Delmastro,
on rare Higgs decays/production by Lindsey Gray, on SUSY and other exotics searches by Monica Weilers, on dark matter searches by Alex Tapper, on 
long lived particles by Juliette Alimena and Marco Drewes,  on multiboson models by  Tiesheng Dai,
on vector-boson scattering by  Narei Lorenzo Martinez, on vector bosons produced in vector boson fusion by Meng Lu, on processes involving W/Z/top  at  LHCb by Oscar De Aguiar, on
precision electroweak measurements by  Elena Yatsenko and Carlos Erice Cid, on the W mass by  Mauro Chiesa, on precision measurements of W/Z jets by  Ewelina Lobodzinska, on 
electroweak corrections to  the Higgs by Armin Schweitzer...).  The conclusion is that so far the LHC has found the Higgs  boson and determined its properties to be consistent with the SM, but has found no BSM physics.

\begin{figure}
\begin{minipage}{0.45\linewidth}
\centerline{\includegraphics[width=0.95\linewidth]{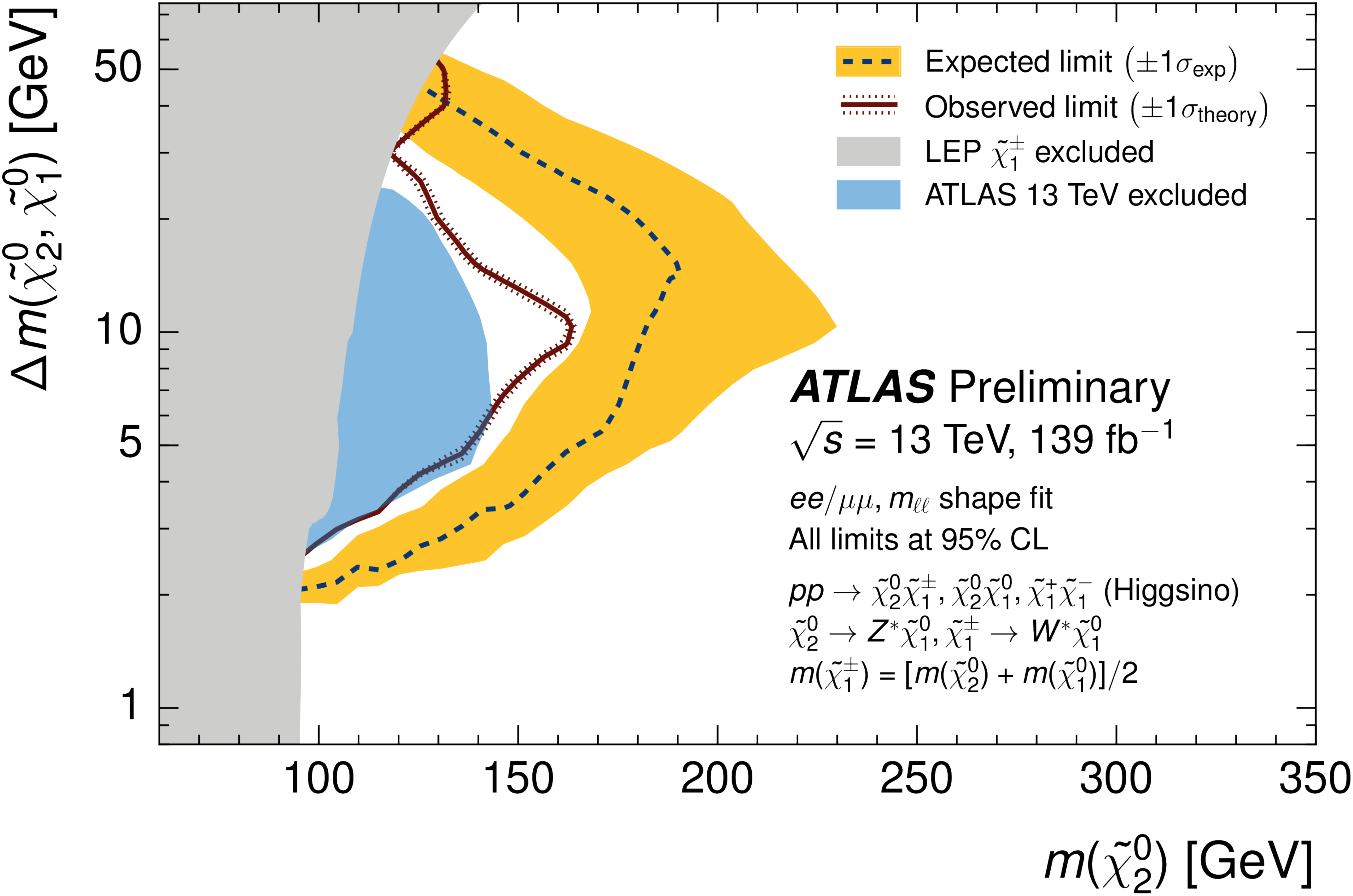}}
\end{minipage}
\hfill
\begin{minipage}{0.53\linewidth}
\centerline{\includegraphics[width=0.95\linewidth]{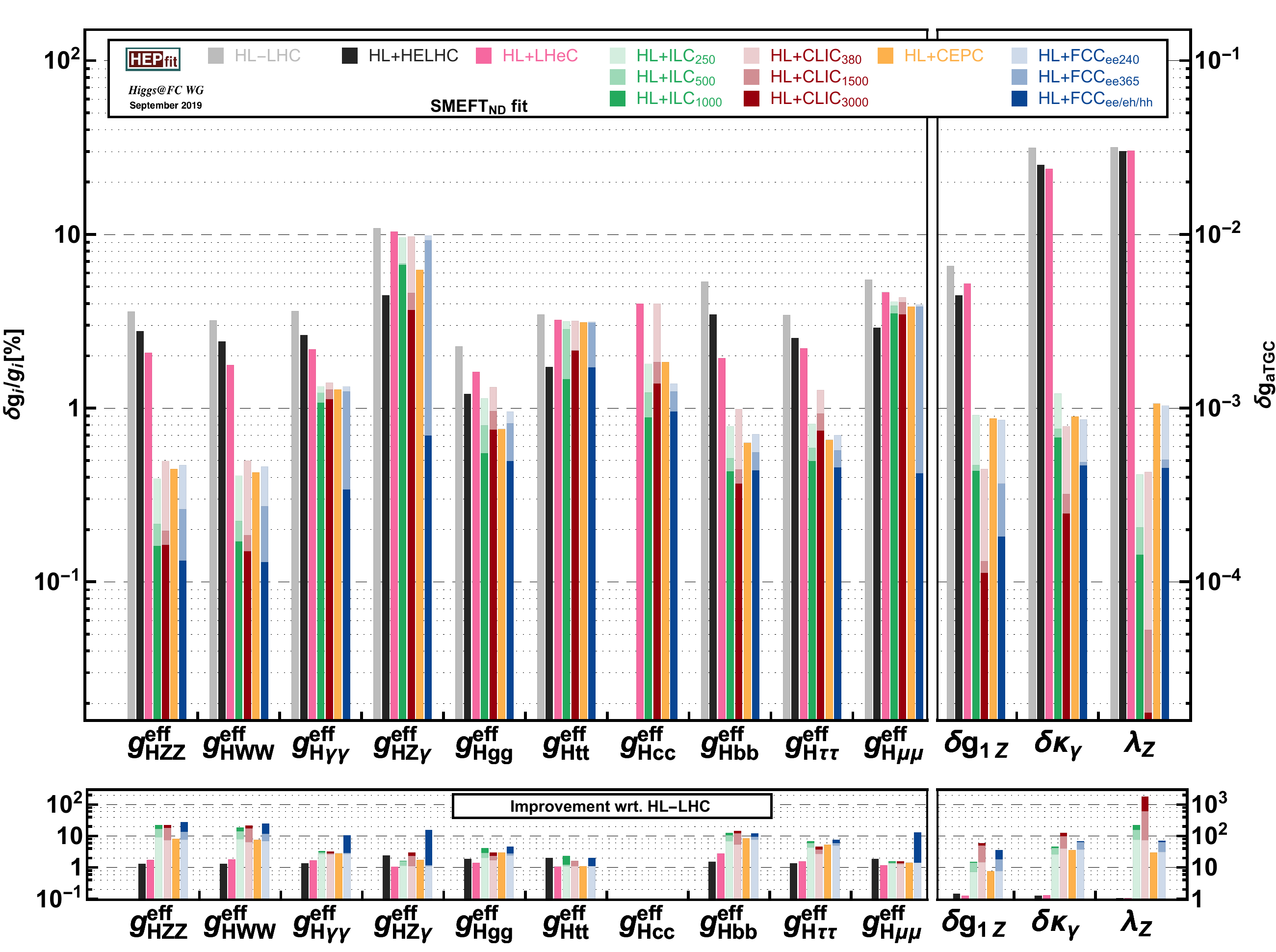}}
\end{minipage}
\caption[]{Left: Fig.~12 of the ATLAS-CONF-2019-014 report reproduced in M. Kado's talk at this conference,  showing the expected 95\% CL exclusion sensitivity (blue dashed line) with $\pm  1 \sigma$ (yellow band) from experimental uncertainties, and observed limit (red solid line), derived from dilepton searches, in the LSP mass $m$ -neutralino mass difference $\Delta m$ plane,   for a compressed neutralino spectrum.  Right: Fig.~3 of~\cite{deBlas:2019rxi} presented in H. Gray's talk,  showing the reach of precision measurements of the Higgs boson couplings, with $\delta g/g$ given in percentage, at the HL-LHC and future collider projects. }
\label{fig:1}
\end{figure}

After the very large number of searches, in a large variety of topologies and models,  the question that arises naturally is if SUSY or other BSM physics could be found in the future at the LHC. The answer is yes, although the emphasis in now on models that ``hide" BSM physics at the LHC, such as in ``Natural SUSY" (see e.g.~\cite{Papucci:2011wy}) or  SUSY with highly compressed spectrum scenarios (see e.g.~\cite{ArkaniHamed:2006mb, Martin:2007gf}). These scenarios would escape detection because signals would be close or under  energy detection  thresholds at the LHC. 
 For example,  the reason  charginos and neutralinos may not have been observed so far may be that their masses are ``compressed", i.e. are all  very close to the mass of the lightest supersymmetric particle (LSP) into which they decay. In this case the decay would only produce pairs of electrons or muons with very low momenta, which are very difficult to identify at the LHC. As an example of a dilepton search M.  Kado  showed  the upper panel of Fig. 12 of the ATLAS-CONF-2019-014 report, reproduced here in the left panel of Fig.~\ref{fig:1}.  This plot shows that  the observed upper limit placed on the mass difference between the two lightest neutralinos given as function of the mass of one of them is worse than the expected limit. Remarkably, the ATLAS sensitivity reaches a few GeV for the mass difference, for neutralino masses between 100 and 150 GeV. The observed limit (red line) is below the 1$\sigma$ (yellow) exclusion sensitivity band, showing that there could be a  signal in the data at more than the 1$\sigma$ level or that it could be one of the many 1 and 2$\sigma$ fluctuations  that appear in data where there is no signal. If with more data the discrepancy between the observed and expected limits would become more significant it could, at some point, lead to the discovery of BSM physics in a channel like this one.

The direct detection searches  at the HL-LHC will reach a discovery potential of  up to O(2-3 TeV) in mass for gluinos, of O(1.5 TeV) for stops and  O(1 TeV) for weakly interacting SUSY particles, of up to 6 TeV for extra Z bosons and up to  8 TeV for extra W bosons, and of 1.5 -2 TeV for leptoquarks (see e.g. M. Kado's review).

If direct searches for new physics fail, there is still the powerful tool  of precision measurements of the Higgs couplings to check if they show any deviation from those predicted by the SM. In order to understand the energy scale reach of  precision measurements we can use a simple ballpark argument.  If the spectrum of the Higgs sector contains one Higgs boson of mass $m_h$ and other heavier scalars of mass  $M$ or larger, then in the decoupling regime in which $M >> m_h$, the influence of these heavy particles on the properties of the Higgs boson is proportional to $m_h^2/M^2$~\cite{Haber:1989xc, Haber:2013mia}. This result does not hold in general for other types of heavy particles with different couplings, but we can take it as a simple estimate.  Using this result, the effects of new physics at the TeV scale on the properties of the Higgs are at the \% level, and those of new physics at the 10 TeV scale are at the 10$^{-4}$ level. This latter level will be difficult to get even at future colliders (see right panel of Fig.~\ref{fig:1}).

Precision measurements of Higgs decays  are at the O(10\%) level at the LHC,  and will be at the O(1\%) level at the HL-LHC (see e.g. M. Kado's talk).  In their talks at the conference about future colliders Alain Blondel and Heather Gray showed that precision measurements of  some Higgs couplings could reach  the O(0.1\%) level at future colliders, such as the High-Energy LHC (HE-LHC)  and the Future Circular Collider (FCC)~\cite{deBlas:2019rxi} (see the right panel of Fig.~\ref{fig:1}). 

The FCC would occupy a new 100 km long tunnel located at CERN and have new 16 T magnets. It could reach a center of mass energy of 100 TeV as hadron collider (FCC-hh) and  of O(100 GeV) as e$^+$e$^-$ collider (FCC-ee). The HE-LHC would reuse the existing LHC tunnel for a hadron collider with an increased  magnetic field, due to 16 T magnets. It  could reach 27 TeV of center of mass energy, with three times the luminosity of the HL-LHC.

A precision of 10$^{-3}$ in the Higgs couplings may not test the 10 TeV scale by itself, but combined with  electroweak precision measurements in a lepton FCC collider, FCC-ee,  BSM energy scales as high as O(10 TeV) could be tested~\cite{Blondel:2019yqr}.  Thus, precision measurements of the Higgs properties at the HL-LHC and future colliders provide a path to discover BSM physics alternative to direct searches.  Alain Blondel expressed it succinctly in his talk: the next facility must be versatile with as broad and powerful reach as possible, as there is no precise target. 

\subsection{Before and after the Planck CMB data}

The Cosmic Background Radiation (CMB) is a remnant from recombination, the moment in the evolution of the Universe when atoms became stable. Just before recombination, the visible matter consisted of a plasma, with separate nuclei and electrons forming a baryon-electron-photon fluid. There were standing pressure waves in this fluid, established when baryons  started falling into potential wells due to dark matter, which increased the temperature of the fluid and thus the  photon pressure, until pressure won  over gravity and produced the expansion of the fluid. This expansion ceased when, as the expanding fluid became colder, the photon pressure decreased and gravity won over again, after which the cycle repeated.  When atoms became stable, photons escaped, and reach us now as the CMB radiation showing the hotter and colder regions as CMB anisotropies, and baryons caught in an expanding phase of the oscillations remained in spherical shells of predictable radius, which are seen as Baryon Acoustic Oscillations (BAO) found in the clustering of galaxies.

\begin{figure}
\begin{minipage}{0.57\linewidth}
\centerline{\includegraphics[width=0.95\linewidth]{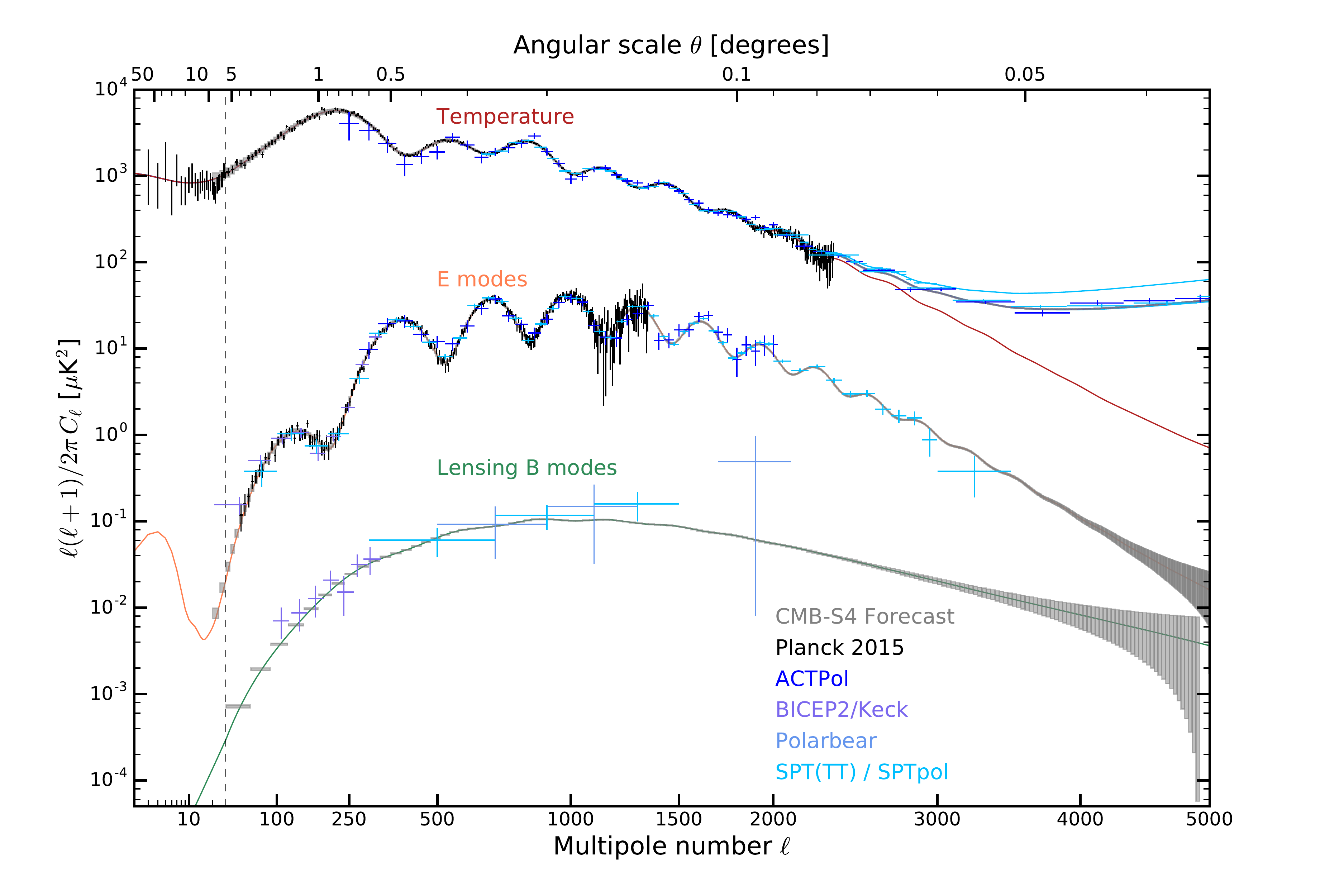}}
\end{minipage}
\hfill
\begin{minipage}{0.41\linewidth}
\centerline{\includegraphics[width=0.95\linewidth]{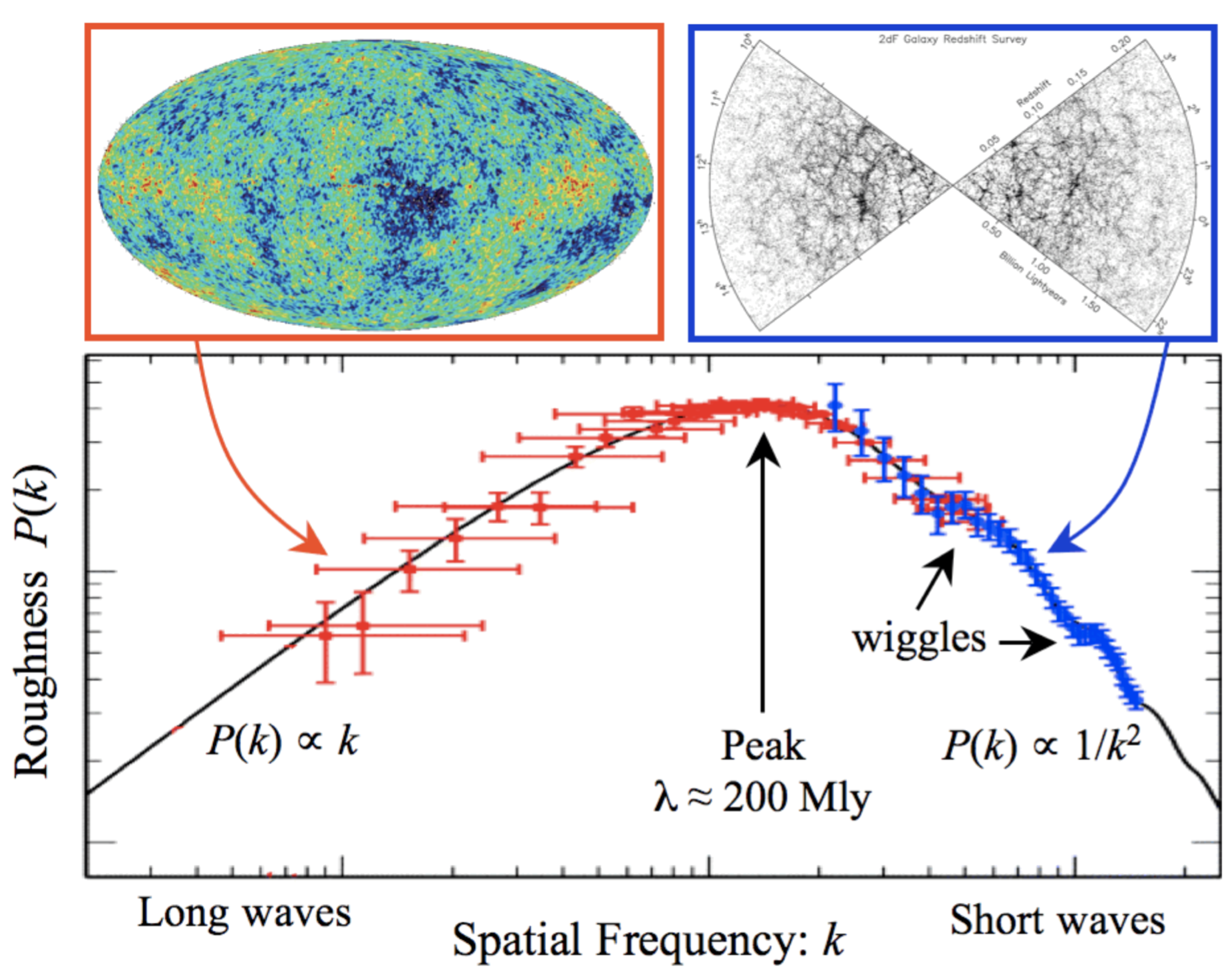}}
\end{minipage}
\caption[]{Figures taken from G. Smoot's talk at the conference. Left: Compilation of current and proposed CMB anisotropy angular power spectrum observations as of 2016 (Fig. 1 of~\cite{Abazajian:2016yjj}).  Right: BAO ``wiggles" in the matter power spectrum~\cite{Anderson:2013zyy}.}
\label{fig:3}
\end{figure}

Before the 2013 Planck precision data~\cite{Ade:2013zuv} (later followed by 2015 and 2018~\cite{Aghanim:2018eyx} data releases) only three of the peaks of the CMB anisotropies angular power spectrum had been measured. After, there  were seven peaks precisely measured in the temperature (Planck TT)  angular power spectrum, besides   measurements of E-mode polarization anisotropies (Planck TE and EE angular power spectra), and other precision measurements (see G. Smoot's talk at this conference and the left panel of Fig.~\ref{fig:3}). The BAO in the matter power spectrum (see G. Smoot's talk  and the right panel of Fig.~\ref{fig:3}) were first detected in  2005,  and measured with higher precision by the  SDSS-III Baryon Oscillation Spectroscopic Survey (BOSS) in 2012 and 2013~\cite{Anderson:2013zyy}.

All CMB and BAO data are consistent with the standard $\Lambda$CDM model, solidifying the interpretation of all available data in terms of a  cosmology based on General Relativity and a content of the Universe consisting of somewhat less than 5\% of  visible matter, described by the SM, about 25\% of dark matter and the remaining about 70\% in dark energy.

An important  consequence of these precision data is that no proposed ``alternative to dark matter"  explains the CMB and BAO spectra after 2013 (see e.g. the talk  of Dan Hooper, ``In Defense of Dark Matter" in a debate with Eric Verlinde at the KITP, Santa Barbara, on 4/30/2018, available at the KITP website). These spectra depend crucially on the depth of the dark matter wells  increasing due to the effect of gravity, while the standing pressure oscillations of the baryon-electron-photon fluid are taking place. After the 2013 precision data, there have been no successful fits to the CMB and BAO data based on theories which attempt to replace dark matter by a modified covariant theory of gravity. E.g. a covariant theory of MOND fit to the first three peaks of the CMB angular spectrum~\cite{Skordis:2005xk} was marginally consistent with data in 2005, but cannot accommodate current CMB data.  In the standard cosmology the  amplitude of the BAO oscillations in the matter power spectrum are small as measured because they are suppressed  as baryons fall into the growing potential wells formed by dark matter, leaving only percent level traces of the primordial oscillations.  Without dark matter the BAO oscillations should be just as apparent in matter as they are in the radiation, thus the amplitude of the oscillations would be  about 30 times larger~\cite{Dodelson:2011qv}.  Thus a test of the validity of any model proposing an alternative to dark matter is  its prediction of the matter power spectrum, not only its predictions for the rotation curves of galaxies, or its description of the dynamics of galaxy clusters.

\subsection{Before and after the LIGO-VIRGO detection of gravity waves}

Before the first direct detection of a binary Black Hole (BH) merger in 2016~\cite{Abbott:2016blz},  the merger of a 36 M$_\odot$ BH and a 29 M$_\odot$ BH to form a 62 M$_\odot$ BH,  gravitational radiation had been detected only indirectly, though the slowdown of  the rotation of pulsars, and there have been no BH's observed with mass of O(10 M$_\odot$) (M$_\odot$ is the solar mass).

The detection of gravity waves gave us  entirely new eyes to observe the Universe. Many $\sim$10 M$_\odot$ BH-BH mergers have been observed, and in 2017 there was the remarkable first multi-messenger observation of a binary neutron-star merger (the GW17081 event)~\cite{GBM:2017lvd}, detected first via gravitational waves,  in which 70 observatories (LIGO, Virgo, Fermi GBM, INTEGRAL, DES, etc) participated, and was  searched but not found in neutrinos by ANTARES, IceCube and Pierre Auger (at the conference this event was mentioned by  Imre Bartos, A. Carnero for the DES collaboration and D. Boncioli for the Auger collaboration).

\begin{figure}
\centerline{\includegraphics[width=0.70\linewidth]{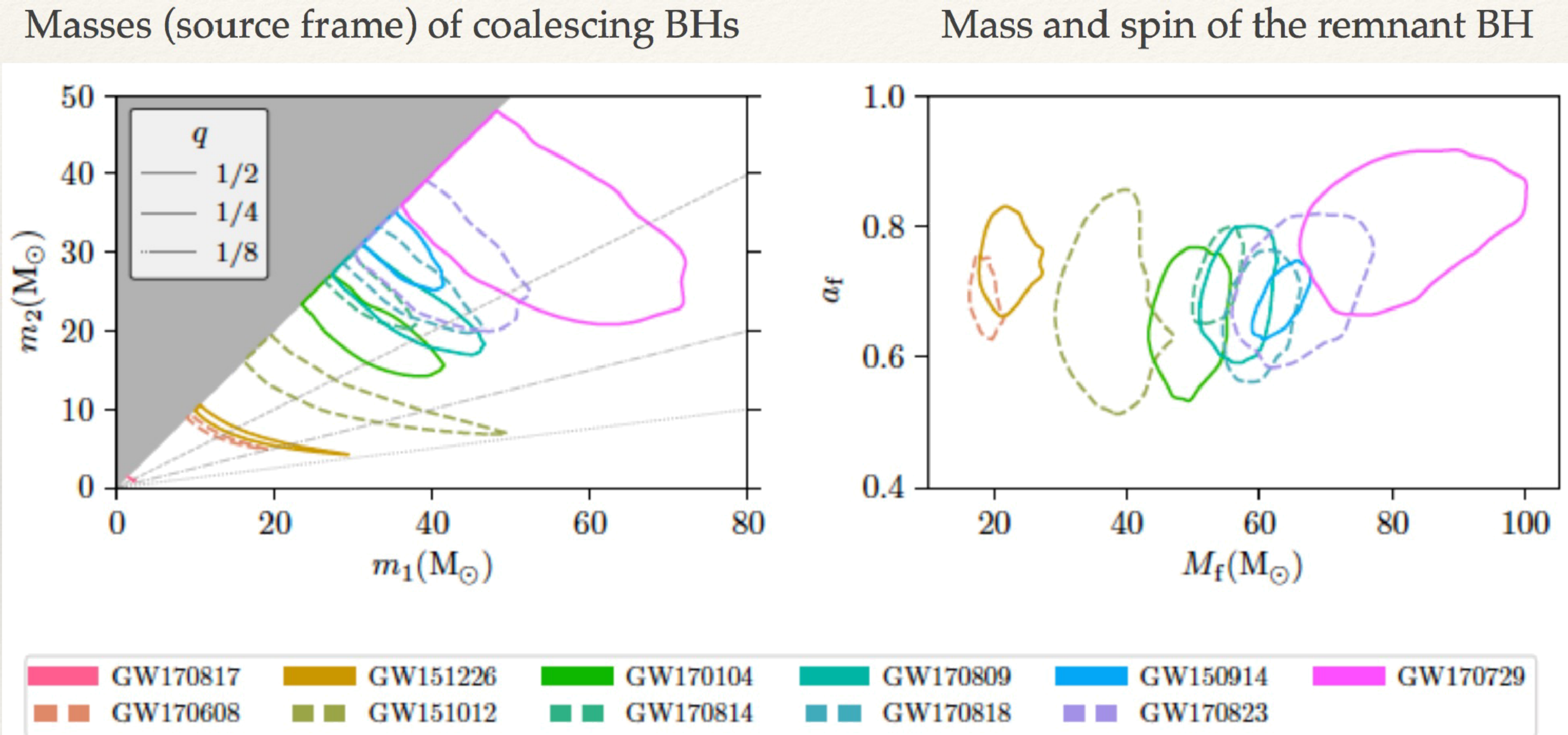}}
\caption[]{Figure reproduced from the talk of  S. Babak at this conference. Mass of merging BHs (left) and mass and spin of  remnant BHs (right) observed by LIGO-VIRGO.
}
\label{fig:2}
\end{figure}

The BH's observed by LIGO and VIRGO can be stellar in origin, i.e. formed by the gravitational collapse of  very massive stars, after they exhaust their nuclear fuel. Only stellar BH with mass M $<$ 70 M$_\odot$, can be formed from a single progenitor, otherwise this progenitor star is unstable. Fig.~\ref{fig:2} shows that most observed merging BH have a mass below 70 M$_\odot$. Larger stellar BH could be formed via mergers of smaller ones.    But,  it has also been suggested that the observed merging BH's could be primordial, instead of stellar, in origin so that these BH could be part of the dark matter~\cite{Bird:2016dcv, Clesse:2016vqa}. Primordial Black Holes (PBH's) are a hypothetical type of BH formed in an early phase transition, before Big Bang Nucleosynthesis (BBN) thus non-baryonic, proposed by  ZelÕdovich and Novikov~\cite{Zeldovich-1996} in 1966, Hawking~\cite{Hawking:1971ei} in 1971 and  Carr and Hawking~\cite{Carr:1974nx} in 1974.

The idea that the LIGO-VIRGO BH's could be PBH's  led to a resurgence of interest in these potential constituents of the dark matter and the recent re-examination of all the observational  limits on  their abundance. As can be seen e.g. in the  recently revised  Fig. 5 of~\cite{Niikura:2017zjd}, in the mass range  10$^{-15}$ to 10$^{-11}$ M$_\odot$  PBH's could constitute the whole of the dark matter, while for all other mass ranges PBH's could only constitute the whole of the dark matter if they are created with a wide distribution of masses (so that  PBH's of a single particular  mass constitute only a small fraction of the DM). BH with mass smaller than 10$^{-15}$ M$_\odot$ would have evaporated by now (and an upper limit on  the abundance of smaller BH's using Voyager data was discussed by Marco Cirelli at this conference).

\section{Anomalies}

According to the Cambridge English Dictionary, an anomaly is ``a person or thing that is different from what is usual, or not in agreement with something else and therefore not satisfactory" and something anomalous is ``deviating from what is standard, normal, or expected."

Since what seems to be expected is that anomalies are not satisfactory, it is to be noted that
in our field we have an anomalous appreciation for anomalies. We love them, as a way to challenge our standard models and make progress. Several anomalies in cosmology, particle physics and astro particle physics were discussed at the conference.

\subsection{Anomaly in cosmology: the value of the cosmological constant}

This anomaly was discussed at the conference by Cristian Rusu for the H0LiCOW collaboration, Aurelio Carnero for the DES collaboration and  Ed Copeland. 

The Hubble constant, H$_0$ can be measured locally and its value can be derived  early Universe observations and there is a tension between measurements of H$_0$~\cite{Riess:2016jrr, Riess:2019cxk} Multiple independent  measurements  in the local Universe (e.g. from Cepheids and Type-Ia supernovae), which are nearly independent of the cosmological expansion history, find $H_0$ = 74.03 $\pm$1.42 km s$^{-1}$ Mpc$^{-1}$~\cite{Riess:2019cxk}. The value inferred from  Planck  CMB data and  from BAO data, assuming the expansion history since recombination in the standard $\Lambda$CDM cosmological model,  is $H_0$ = 67.66 $\pm$ 0.42 km s$^{- 1}$ Mpc$^{-1}$~\cite{Aghanim:2018eyx}. The discrepancy is significant at $\sim 4.4 \sigma$ level (Cristian Rusu  announced at the conference that including a recent measurement of the H0LiCOW collaboration the significance level is more than 5$\sigma$~\cite{Wong:2019kwg})

Many new physics solutions have been proposed to alleviate this tension~\cite{Riess:2019cxk},  either modifying cosmology at the moment of CMB emission so that the CMB/BAO inferred value increases, or modifying the cosmology at late-times, so that the expansion rate matches the CMB data at decoupling and the local rate today (e.g. \cite{DiValentino:2017oaw, Mortsell:2018mfj, Poulin:2018zxs, Vattis:2019efj, Gelmini:2019deq})

\subsection{Anomalies in particle physics: in $B$ meson decays}

These anomalies were discussed  by  Marumi Kado, Mitesh Patel and Tobias Hurth at the conference. In recent years, several deviations from SM prediction have been building up in B-decay measurements (see the talk of M. Patel).  Discrepancies were found in  the following ratios of decay branching ratios ${\mathcal{B}}$ of  $B$ and $\bar{B}$ decays into $D$ and $D^*$ and into $K$ and $K^*$ mesons:

\[R(D^{(*)}) =\frac{{\mathcal{B}}(\bar{B} \to D^{(*)} \tau^- \bar \nu_\tau)}{{\mathcal{B}}(\bar{B} \to D^{(*)} \ell^- \bar \nu_\ell)} ~~~~{\rm and}~~~  
R(K^{(*)}) =\frac{{\mathcal{B}}(B \to K^{(*)} \mu^+ \mu^-)}{{\mathcal{B}}( B \to K^{(*)} e^+ e^-)} ~.         \]

 The discrepancies in the $R(D)$ and $R(D^*)$ ratios were at  the 4.4$\sigma$ level before new 2019  Belle results~\cite{Prim:2019wem} reduced them to  the 3.1$\sigma$ level.  The $R(K)$ and $R(K^*)$ ratios, predicted to be  $R(K) = R(K^*) = 1$ in  the SM,  after the LHCb run 1  were found to be   $\simeq$ 0.7, a discrepancy at the 2.5$\sigma$ level~\cite{Aaij:2014ora}.  In the LHCb run 2, the ratios were closer to the SM, $\simeq$ 0.9~\cite{Aaij:2019wad}. However, since the combined uncertainty of both runs is smaller, the discrepancy with the SM of the combined data of both runs still shows a significance at the same 2.5$\sigma$ level from the SM~\cite{Aaij:2019wad}.   Also the B$_{s,d} \to  \mu^+ \mu^-$ branching ratio measured by ATLAS in 2018~\cite{Aaboud:2018mst}, combined with the previous 2013-2014 CMS  and LHCb measurements~\cite{CMS:2014xfa} show a small tension at the 2$\sigma$ level with the  SM prediction~\cite{Aebischer:2019mlg}.
 
Models exist to solve these anomalies separately or simultaneously (see  the talk of T. Hurth and e.g.~\cite{Aebischer:2019mlg} and references therein), but  new data seems to move these anomalies back towards  the SM, so it is not clear if any of them will persist to become a signature of BSM physics. There is a rich B physics  program ahead with future new data from LHCb and Belle II which will confirm or reject them.

\subsection{Anomaly in particle physics: LSND/MiniBooNE $\bar\nu_e$ excess}

 The long-standing excesses observed in the $\bar\nu_\mu \to  \bar\nu_e$ appearance channel in the short-baseline LSND~\cite{Aguilar:2001ty} and MiniBooNE~\cite{Aguilar-Arevalo:2013pmq}  experiments have been recently strengthened with additional 2018 MiniBooNE data~\cite{Aguilar-Arevalo:2018gpe}. Their interpretation in terms of a 4th, sterile, neutrino  requires this neutrino to have a mass close to 1eV and  large couplings $U_{e 4}$ and $U_{\mu 4}$ to both $\nu_e$ and $\nu_\mu$, so that the $\nu_\mu$-$\nu_e$ mixing is   $\sin^2 2\theta_{\mu e} = 4~|U_{\mu 4}|^2~|U_{e 4}|^2$.  This interpretation is in strong, $4.7\sigma$, tension (see e.g.~\cite{Dentler:2018sju}) with $\nu_\mu$ disappearance data from IceCube~\cite{TheIceCube:2016oqi}  and MINOS+~\cite{Adamson:2017uda}, which test $|U_{e 4}|^2$.
 
This was discussed in parallel session by Carlos Arg\"uelles, who presented explanations  alternative to a sterile neutrino. Since MiniBooNE  cannot  distinguish between e and $\gamma$, the emission of a  $\gamma$  or an exotic $Z'$ emission may play a role in its anomalous results and some unexpected nuclear physics effect or background could explain the LSND results.

\subsection{Anomaly in astroparticles  physics: 3.5 keV X-ray line - a 7 keV sterile neutrino?}

A 3.5 keV X-ray  emission line was found in 2014 in 74 stacked images of galaxy clusters~\cite{Bulbul:2014sua}  and coming
from the Andromeda galaxy and  the Perseus cluster~\cite{Boyarsky:2014jta}, which remains a subject of lively debate (see e.g.~\cite{Dessert:2018qih} and  \cite{Boyarsky:2018ktr}). It  could be due to the radiative decay $\nu_s \to \nu_a \gamma$ of sterile neutrinos comprising all or part of the dark matter, with mass $m_s=7$ keV since  $E_\gamma = m_s/2$.
 
Assuming they account for all of the dark matter, the signal could correspond to sterile neutrinos produced via resonant active-sterile oscillations, which would require a large lepton asymmetry L $\simeq 5 \times10^{-4}$~(see e.g.~\cite{Abazajian:2017tcc}). In this case its mixing angle  should be $\sin^2 2\theta \simeq 10^{-10}$.  However, these neutrinos would be produced before BBN, a yet unexplored epoch (the earliest cosmological remnant detected so far are the light nuclei produced in BBN). Thus assumptions about cosmology must be made to compute their relic abundance.
Alternative assumptions, also compatible with all observations, may lead to explain the signal with a very different mixing, e.g.
$\sin^2 2\theta \simeq 10^{-7}$~\cite{Abazajian:2017tcc} in a low reheating temperature model~\cite{Gelmini:2004ah}, a mixing which is within reach of the KATRIN tritium decay experiment~\cite{Mertens:2018vuu} (in this case, sterile neutrinos would account for only a fraction 0.7 $\times 10^{-3}$ of the dark matter). 

The ESA's XMM-Newton  and  NASA's Chandra  satellites used to find the line do not provide enough energy resolution of the line to test its origin.  JAXA's ASTRO-H (called Hitomi after  its  first light),  launched on Feb. 17, 2016, was expected to measure the profile of the line and prove or disprove that it is due to dark matter in 1 yr of data taking. Unfortunately this satellite iwas destroyed on March 26, 2016.  The best prospect to test this line resides now in the
 JAXA-NASA ``X-Ray Astronomy Recovery Mission" (XARM) project  which will launch in  2021. The next planned X-ray astronomy satellite is ESA's ATHENA, but it is scheduled for  2028.

\subsection{Anomaly in astroparticle physics: the PAMELA/AMS 10-100's GeV positron excess}

 This was discussed at the conference  by Marco Cirelli,  and by Harm Schoorlemmer for the HAWC collaboration.  The $e^+$ excess flux in the 10 GeV to 100's of GeV energy range with respect to what is expected from secondary cosmic rays was found by PAMELA~\cite{Adriani:2010rc} in 2008 (and even earlier by the HEAT balloon experiment~\cite{Barwick:1997ig}) and then confirmed by the Fermi Space Telescope-LAT~\cite{FermiLAT:2011ab}  and  the AMS~\cite{Aguilar:2013qda} collaborations.
 
 In 2017, HAWC's first measurements~\cite{Abeysekara:2017hyn} of the very high-energy (multi-TeV) $\gamma$-ray emission from the Geminga and Monogem pulsars show that they inject a flux of positrons into the local interstellar medium large enough to account for the observed $e^+$ excess. This strongly favors nearby pulsars as the origin of this signal~\cite{Hooper:2017gtd}, instead of dark matter annihilation.

\subsection{Anomaly in astroparticle physics: the  GeV Galactic Center excess}

This anomaly is discussed  in the posted slides of Marco Cirelli's talk at the conference. Subtracting from the Fermi  LAT data all known contributions, an excess of GeV photons coming from the Galactic Center and the Inner Galaxy has been found by a variety of independent studies, since the initial work of D. Hooper and collaborators in 2009-2010~\cite{Goodenough:2009gk, Hooper:2010mq, Hooper:2011ti}.     The existence of this excess has been confirmed, but its origin is still uncertain. The most pursued explanations are either annihilating or decaying dark matter or unresolved astrophysical sources,  among which prime candidates are millisecond pulsars~\cite{Abazajian:2010zy, Bartels:2015aea, Lee:2015fea}. Known millisecond pulsars have an emission spectrum which reproduces well the observed spectrum of the excess, however discussions of the viability of this interpretation center on the existence and the potential origin of the necessary large population of millisecond pulsars.    

A tentative evidence  of the existence of an unresolved point source population was found in recent years using the statistics of the photons in the GeV excess flux (see e.g.~\cite{Bartels:2015aea}) which seemed to have disfavored dark matter as the origin of the signal. Besides,  it was claimed~\cite{Macias:2019omb} that stars associated with the galactic bulge, a few kpc triaxial bar-like structure in the central region of our Galaxy, also could contribute significantly to the GeV excess, leaving no much room for dark matter.  However, a very recent study~\cite{Leane:2019xiy} from April 2019 (with the suggestive title 
``Dark Matter Strikes Back at the Galactic Center") found that due to mis-modeling effects in the real Fermi data, large artificial injected dark matter signals can be completely misattributed to point sources and, consequently, that dark matter may provide a dominant contribution to the GeV Galactic Center excess after all.   

In any event, given the large astrophysical uncertainties, this could possibly constitute at most a corroboration of a discovery of dark matter somewhere else.

\subsection{Anomaly in cosmology: the EDGES 21 cm signal}

This anomaly was discussed in a parallel session talk by Ely Kovetz, concentrating on millichaged dark matter~\cite{Kovetz:2018zan}. EDGES (the Experiment to Detect the Global Epoch of Reionization Signature), which measures the global 21cm line emission from neutral hydrogen,  announced in 2018
the first detection of a signal from the ``cosmic dawn", the epoch in the Universe when stars first appeared. The signal was an absorption feature due to gas at a redshift of about 17,  much deeper than expected in the standard cosmological model.  The signal could be explained if the gas producing the absorption line is much colder  than the CMB than predicted by the standard cosmology. Thus, some possible explanations are that interactions with some dark matter component (such as millicharged dark matter) cooled the gas,  or that dark matter decays heated up the background radiation.  

Alternatively, the feature could actually be a not yet understood artifact of an imperfect background subtraction.  In this regard,  there are several  future 21 cm observatories, presented in the overview of Miguel  Floranes at the conference, such as   LEDA, PRIZM, SARAS2, HERA and SKA, which will be able to confirm or reject the EDGES feature.

  \subsection{Anomaly in astropaticle physics: the DAMA/LIBRA  annual modulation signal}
  
  For the past 20 years the DAMA/LIBRA NaI(Tl) direct dark matter detection experiment has found an annual modulation
  in its data  compatible with the modulation expected for a dark matter signal, a maximum of the interaction rate at the beginning of June and a minimum six months later (coinciding with the times in the year when the velocity of Earth with respect to the Galaxy is maximum and  minimum).  With  a very large exposure of  2.46 ton yr, DAMA/LIBRA finds a 12.9$\sigma$ C.L. modulation signal  in 2-6 keVee (keV electron-equivalent) recoil energy range   (a 9.5$\sigma$  modulation  signal in  the  1-6 keVee energy range)~\cite{Bernabei:2018yyw}. However, this signal remains incompatible with the results of other direct detection experiments for all the many dark matter candidates  that have been studied so far.   It is clear that the experiment sees an annual modulation, the issue is to determine what is that it is observing. It seems clear that the effect should be found
by another similar experiment to elucidate its origin.  
  
For the first time a recently established ``Global  NaI(Tl) Collaborative Effort" will check this modulation signal using the same target material of DAMA. This is a collaboration between COSINE-100   (combining the previous DM-Ice and KIMS experiments, which have 55 kg and 52 of NaI,  respectively)  located in  the YangYang Laboratory in S. Korea, ANAIS (with 112 kg) in the  Canfranc Laboratory in Spain, and SABRE (with 50 kg) in two sites, the Gran Sasso Laboratory in Italy and the Stawell Laboratory in Australia. The latter is the only underground laboratory in the Southern Hemisphere. It is important to check a potential annual modulation in the signal in both hemispheres, because any annually modulated background would be correlated with the maximum and minimum temperature of the upper atmosphere, would be maximum in summer and minimum in winter, which is approximately the phase expected of a dark matter signal in the Northern Hemisphere, but would be entirely out of phase in the Southern Hemisphere.

 Results presented at the conference  for the ANAIS collaboration  by Maria Martinez  reject an annual modulation at the $1.8\sigma$ level, and more statistics is  needed  to make a significant statement (a 3$\sigma$ level will be reached in 2.5 yr of data taking). Results of the  COSINE-100 collaboration presented at the conference  by Kyungwon Kim  show no modulation, but they are also statistically limited still. 
  
  There are many other direct dark matter detection efforts worldwide. Present, upcoming and proposed direct detection experiments searching for dark matter particles with mass of O(GeV) or larger, the traditional WIMP mass range, such as XENON1T,  LZ,  XENONnT,  DarkSide20T, SuperCDMS, PICO, DARWIN, COSINE-100, ANAIS, SABRE etc  and directional detectors, were presented at the conference in the overview of Priscilla Cushman and parallel session talks by J.P. Zopounidis, Dimitri Missiac, Patricia Sanchez-Lucas and  Murat Ali G\"uler, among others.   
 
Many ideas for the direct detection of sub-GeV mass ``Light Dark  Matter"  (LDM)  that are being actively explored (see e.g. \cite{Alexander:2016aln}, \cite{Battaglieri:2017aum}) were presented at the conference in the overview of Rouven Essig. We heard specifically about two of them, SENSEI and  ABRACADABRA.

SENSEI~\cite{Abramoff:2019dfb}, presented by Rouven Essig, uses silicon Skipper-CCDs consisting of O(million) pixels each.  A prototype consisting of a single Skipper-CCD  in the MINOS cavern at the FNAL  set limits with an exposure of 0.177 gram-days on dark matter-electron scattering for masses  of 500 keV to 5 MeV, and on dark-photon dark matter being absorbed by electrons for masses under 12.4 eV. A 100 g version of SENSEI has been funded and will start in 2020 at SNOLAB.  

ABRACADABRA is based on a new idea to detect  sub-$\mu$eV axions and axion-like particles via their interaction with the azimuthal magnetic field of a  toroidal magnet,  which sources an effective current parallel to the magnet. The first results from ABRACADABRA-10 cm~\cite{Salemi:2019xgl} were presented by Jonathan Ouellet at the conference.

\section{Neutrinos}

Let me finally mention two ideas which I believe worth singling out for their innovative character. They both refer to neutrino physics. This is a  very active field of research, involving vastly different  subfields, from  high-energy neutrino detection, to sterile neutrino searches, to 0$\nu \beta \beta$ decay searches etc. In the conference we heard e.g. from Luigi Fusco about ORCA, Stefan Schoppmann about  STEREO, Ann Schutz about GERDA,  Guido Fantini about CUORE, Luca Gironi about CUPID-0, Claudia Nones about CUPID-Mo and Justo Martin-Albo about NEXT.

One of the ideas deserving special mention was presented by Gwenha\"el de Wasseige. It  is a 
 proposal to detect GeV neutrinos produced in solar flares and other transients with IceCube and KM3NeT, by optimizing the detection window using Fermi LAT $\gamma$-ray data.  The second idea, presented by Sergio Palomares-Ruiz,  is using atmospheric neutrinos passing through Earth observed by IceCube  to do ``Earth Tomography". What is surprising is that existing IceCube data already yield  a rough measurement of the density profile of Earth's interior~\cite{Donini:2018tsg}, and could provide a much better one with a total of 10 yr of data.
 
\section{Prospect} 

Particle physics and cosmology/astrophysics are inextricably intertwined forming a tremendously vibrant, data driven field. I feel very lucky to be in it at this time. 

The Rencontres de Blois are unique in terms of how wide a spectrum of subfields the conference series covers, including theoretical and observational/experimental activities in all of them,  fostering synergy  that is essential to make progress. 

There are many anomalies to resolve and fundamental discoveries to make. There will be many more Rencontres at Blois.

\section*{Acknowledgments}

My thanks go to the organizers of this wonderful conference in  this superb place, the Ch\^ateau Royal de Blois. 
This work was supported in part  by the U.S. Department of Energy (DOE) Grant No. DE-SC0009937.


\section*{References}

\begin{thebibliography}{99}



\bibitem{Espinosa:1998xj} 
  J.~R.~Espinosa and J.~F.~Gunion,
  {\it{ Phys.\ Rev.\ Lett.\ }}  {\bf 82}, 1084 (1999)
  [hep-ph/9807275].
  
\bibitem{Papucci:2011wy} 
  M.~Papucci, J.~T.~Ruderman and A.~Weiler,
  {\it{JHEP }} {\bf 1209}, 035 (2012)
  [arXiv:1110.6926 [hep-ph]].
  
\bibitem{ArkaniHamed:2006mb} 
  N.~Arkani-Hamed, A.~Delgado and G.~F.~Giudice,
  {\it{Nucl.\ Phys.\ }} B {\bf 741}, 108 (2006)
  [hep-ph/0601041].
  
\bibitem{Martin:2007gf} 
  S.~P.~Martin,
  {\it{Phys.\ Rev.\ }} D {\bf 75}, 115005 (2007)
  [hep-ph/0703097 [HEP-PH]].
  
\bibitem{Haber:1989xc} 
  H.~E.~Haber and Y.~Nir,
  {\it{Nucl.\ Phys.\ }} B {\bf 335}, 363 (1990).
 
\bibitem{Haber:2013mia} 
  H.~E.~Haber,
  arXiv:1401.0152 [hep-ph].
  
\bibitem{deBlas:2019rxi} 
  J.~de Blas {\it et al.},
  arXiv:1905.03764 [hep-ph].

\bibitem{Blondel:2019yqr} 
  A.~Blondel {\it et al.},
  arXiv:1906.02693 [hep-ph].
  
\bibitem{Abazajian:2016yjj} 
  K.~N.~Abazajian {\it et al.} [CMB-S4 Collaboration],
  arXiv:1610.02743 [astro-ph.CO].

\bibitem{Anderson:2013zyy} 
  L.~Anderson {\it et al.} [BOSS Collaboration],
  {\it{Mon.\ Not.\ Roy.\ Astron.\ Soc.\ }} {\bf 441}, no. 1, 24 (2014)
  [arXiv:1312.4877 [astro-ph.CO]].
    
\bibitem{Ade:2013zuv} 
  P.~A.~R.~Ade {\it et al.} [Planck Collaboration],
  {\it{Astron.\ Astrophys.\ }}  {\bf 571}, A16 (2014)
  [arXiv:1303.5076 [astro-ph.CO]].
  
\bibitem{Aghanim:2018eyx} 
  N.~Aghanim {\it et al.} [Planck Collaboration],
  arXiv:1807.06209 [astro-ph.CO].
    
\bibitem{Skordis:2005xk} 
  C.~Skordis, D.~F.~Mota, P.~G.~Ferreira and C.~Boehm,
  {\it{Phys.\ Rev.\ Lett.\ }}  {\bf 96}, 011301 (2006)
  [astro-ph/0505519].
  
\bibitem{Dodelson:2011qv} 
  S.~Dodelson,
  {\it{Int.\ J.\ Mod.\ Phys.\ }} D {\bf 20}, 2749 (2011)
  [arXiv:1112.1320 [astro-ph.CO]].
  
\bibitem{Abbott:2016blz} 
  B.~P.~Abbott {\it et al.} [LIGO Scientific and Virgo Collaborations],
  {\it{Phys.\ Rev.\ Lett.\ }} {\bf 116}, no. 6, 061102 (2016)
  [arXiv:1602.03837 [gr-qc]].
  
\bibitem{GBM:2017lvd} 
  B.~P.~Abbott {\it et al.} [LIGO Scientific and Virgo and Fermi GBM and INTEGRAL and IceCube and IPN and Insight-Hxmt and ANTARES and Swift and Dark Energy Camera GW-EM and DES and DLT40 and GRAWITA and Fermi-LAT and ATCA and ASKAP and OzGrav and DWF (Deeper Wider Faster Program) and AST3 and CAASTRO and VINROUGE and MASTER and J-GEM and GROWTH and JAGWAR and CaltechNRAO and TTU-NRAO and NuSTAR and Pan-STARRS and KU and Nordic Optical Telescope and ePESSTO and GROND and Texas Tech University and TOROS and BOOTES and MWA and CALET and IKI-GW Follow-up and H.E.S.S. and LOFAR and LWA and HAWC and Pierre Auger and ALMA and Pi of Sky and DFN and ATLAS Telescopes and High Time Resolution Universe Survey and RIMAS and RATIR and SKA South Africa/MeerKAT Collaborations and AstroSat Cadmium Zinc Telluride Imager Team and AGILE Team and 1M2H Team and Las Cumbres Observatory Group and MAXI Team and TZAC Consortium and SALT Group and Euro VLBI Team and Chandra Team at McGill University],
 {\it{ Astrophys.\ J.\ }} {\bf 848}, no. 2, L12 (2017)
  [arXiv:1710.05833 [astro-ph.HE]]. 
  
\bibitem{Bird:2016dcv} 
  S.~Bird, I.~Cholis, J.~B.~Mu–oz, Y.~Ali-Ha•moud, M.~Kamionkowski, E.~D.~Kovetz, A.~Raccanelli and A.~G.~Riess,
  {\it{Phys.\ Rev.\ Lett.\ }}  {\bf 116}, no. 20, 201301 (2016)
  [arXiv:1603.00464 [astro-ph.CO]].
  
\bibitem{Clesse:2016vqa} 
  S.~Clesse and J.~Garc'a-Bellido,
  {\it{Phys.\ Dark Univ.\ }}  {\bf 15}, 142 (2017)
  [arXiv:1603.05234 [astro-ph.CO]].
  
  \bibitem{Zeldovich-1996}
Y.~B. Zel'dovich and I.~D. Novikov,
{\it{Sov. Astron. }} {\bf 10}, 602 (1967).

\bibitem{Hawking:1971ei} 
  S.~Hawking,
  {\it{Mon.\ Not.\ Roy.\ Astron.\ Soc.\  }} {\bf 152}, 75 (1971).

\bibitem{Carr:1974nx} 
  B.~J.~Carr and S.~W.~Hawking,
  {\it{Mon.\ Not.\ Roy.\ Astron.\ Soc.\ }}  {\bf 168}, 399 (1974).
  
\bibitem{Niikura:2017zjd} 
  H.~Niikura {\it et al.},
  {\it{Nat.\ Astron.\ }}  {\bf 3}, no. 6, 524 (2019)
  [arXiv:1701.02151 [astro-ph.CO]].
  
\bibitem{Riess:2016jrr} 
  A.~G.~Riess {\it et al.},
 {\it{ Astrophys.\ J.\ }}  {\bf 826}, no. 1, 56 (2016)
  [arXiv:1604.01424 [astro-ph.CO]].
  
\bibitem{Riess:2019cxk} 
  A.~G.~Riess, S.~Casertano, W.~Yuan, L.~M.~Macri and D.~Scolnic,
 {\it{Astrophys.\ J.\ }} {\bf 876}, no. 1, 85 (2019)
  [arXiv:1903.07603 [astro-ph.CO]].
  
\bibitem{Wong:2019kwg} 
  K.~C.~Wong {\it et al.},
  arXiv:1907.04869 [astro-ph.CO].
  
  
\bibitem{DiValentino:2017oaw} 
  E.~Di Valentino, C.~B¿ehm, E.~Hivon and F.~R.~Bouchet,
  {\it{Phys.\ Rev.\ }} D {\bf 97}, no. 4, 043513 (2018)
  [arXiv:1710.02559 [astro-ph.CO]].
  
\bibitem{Mortsell:2018mfj} 
  E.~Mšrtsell and S.~Dhawan,
  {\it{JCAP}} {\bf 1809}, 025 (2018)
  [arXiv:1801.07260 [astro-ph.CO]].

\bibitem{Poulin:2018zxs} 
  V.~Poulin, K.~K.~Boddy, S.~Bird and M.~Kamionkowski,
  {\it{Phys.\ Rev.\ }} D {\bf 97}, no. 12, 123504 (2018)
  [arXiv:1803.02474 [astro-ph.CO]].

\bibitem{Vattis:2019efj} 
  K.~Vattis, S.~M.~Koushiappas and A.~Loeb,
  {\it{Phys.\ Rev.\ }} D {\bf 99}, no. 12, 121302 (2019)
  [arXiv:1903.06220 [astro-ph.CO]].

\bibitem{Gelmini:2019deq} 
  G.~B.~Gelmini, A.~Kusenko and V.~Takhistov,
  arXiv:1906.10136 [astro-ph.CO].
  
\bibitem{Prim:2019wem} 
  M.~T.~Prim [Belle II Collaboration],
  arXiv:1906.09337 [hep-ex].

\bibitem{Aaij:2014ora} 
  R.~Aaij {\it et al.} [LHCb Collaboration],
  {\it{Phys.\ Rev.\ Lett.\ }}  {\bf 113}, 151601 (2014)
  [arXiv:1406.6482 [hep-ex]].
  
\bibitem{Aaij:2019wad}
  R.~Aaij {\it et al.} [LHCb Collaboration],
  {\it{Phys.\ Rev.\ Lett.\ }}  {\bf 122} (2019) no.19,  191801
  [arXiv:1903.09252 [hep-ex]]. 
  
\bibitem{Aaboud:2018mst}
  M.~Aaboud {\it et al.} [ATLAS Collaboration],
  {\it{JHEP}} {\bf 1904} (2019) 098
  [arXiv:1812.03017 [hep-ex]]. 
  
  
\bibitem{CMS:2014xfa} 
  V.~Khachatryan {\it et al.} [CMS and LHCb Collaborations],
  {\it{Nature}} {\bf 522}, 68 (2015)
  [arXiv:1411.4413 [hep-ex]].
  
\bibitem{Aebischer:2019mlg} 
  J.~Aebischer, W.~Altmannshofer, D.~Guadagnoli, M.~Reboud, P.~Stangl and D.~M.~Straub,
  arXiv:1903.10434 [hep-ph].
  

\bibitem{Aguilar:2001ty} 
  A.~Aguilar-Arevalo {\it et al.} [LSND Collaboration],
  {\it{Phys.\ Rev.\ }} D {\bf 64}, 112007 (2001)
  [hep-ex/0104049].

\bibitem{Aguilar-Arevalo:2013pmq} 
  A.~A.~Aguilar-Arevalo {\it et al.} [MiniBooNE Collaboration],
  {\it{Phys.\ Rev.\ Lett.\ }}  {\bf 110}, 161801 (2013)
  [arXiv:1303.2588 [hep-ex]].
  
\bibitem{Aguilar-Arevalo:2018gpe} 
  A.~A.~Aguilar-Arevalo {\it et al.} [MiniBooNE Collaboration],
  {\it{Phys.\ Rev.\ Lett.\ }}  {\bf 121}, no. 22, 221801 (2018)
  [arXiv:1805.12028 [hep-ex]].
  
\bibitem{Dentler:2018sju}
  M.~Dentler, ç.~Hern‡ndez-Cabezudo, J.~Kopp, P.~A.~N.~Machado, M.~Maltoni, I.~Martinez-Soler and T.~Schwetz,
  {\it{JHEP}} {\bf 1808} (2018) 010
  [arXiv:1803.10661 [hep-ph]].

\bibitem{TheIceCube:2016oqi} 
  M.~G.~Aartsen {\it et al.} [IceCube Collaboration],
  {\it{Phys.\ Rev.\ Lett.\ }} {\bf 117}, no. 7, 071801 (2016)
  [arXiv:1605.01990 [hep-ex]].
  
\bibitem{Adamson:2017uda} 
  P.~Adamson {\it et al.} [MINOS+ Collaboration],
  {\it{Phys.\ Rev.\ Lett.\ }}  {\bf 122}, no. 9, 091803 (2019)
  [arXiv:1710.06488 [hep-ex]].
  
\bibitem{Bulbul:2014sua} 
  E.~Bulbul, M.~Markevitch, A.~Foster, R.~K.~Smith, M.~Loewenstein and S.~W.~Randall,
  {\it{Astrophys.\ J.\ }}  {\bf 789}, 13 (2014)
  [arXiv:1402.2301 [astro-ph.CO]].
  
\bibitem{Boyarsky:2014jta} 
  A.~Boyarsky, O.~Ruchayskiy, D.~Iakubovskyi and J.~Franse,
  {\it{Phys.\ Rev.\ Lett.\ }}  {\bf 113}, 251301 (2014)
  [arXiv:1402.4119 [astro-ph.CO]].
  
\bibitem{Dessert:2018qih} 
  C.~Dessert, N.~L.~Rodd and B.~R.~Safdi,
  arXiv:1812.06976 [astro-ph.CO].
  
\bibitem{Boyarsky:2018ktr} 
  A.~Boyarsky, D.~Iakubovskyi, O.~Ruchayskiy and D.~Savchenko,
  arXiv:1812.10488 [astro-ph.HE].
  
\bibitem{Abazajian:2017tcc} 
  K.~N.~Abazajian,
  {\it{Phys.\ Rept.\ }} {\bf 711-712}, 1 (2017)
  [arXiv:1705.01837 [hep-ph]].
  
\bibitem{Gelmini:2004ah} 
  G.~Gelmini, S.~Palomares-Ruiz and S.~Pascoli,
  {\it{Phys.\ Rev.\ Lett.\ }}  {\bf 93}, 081302 (2004)
  [astro-ph/0403323].
  
\bibitem{Mertens:2018vuu} 
  S.~Mertens {\it et al.} [KATRIN Collaboration],
  {\it{J.\ Phys.\ }} G {\bf 46}, no. 6, 065203 (2019)
  [arXiv:1810.06711 [physics.ins-det]].

\bibitem{Adriani:2010rc} 
  O.~Adriani {\it et al.} [PAMELA Collaboration],
  {\it{Phys.\ Rev.\ Lett.\ }}  {\bf 105}, 121101 (2010)
  [arXiv:1007.0821 [astro-ph.HE]].
 
\bibitem{Barwick:1997ig} 
  S.~W.~Barwick {\it et al.} [HEAT Collaboration],
  {\it{Astrophys.\ J.\ }} {\bf 482}, L191 (1997)
  [astro-ph/9703192].
  
\bibitem{FermiLAT:2011ab} 
  M.~Ackermann {\it et al.} [Fermi-LAT Collaboration],
  {\it{Phys.\ Rev.\ Lett.\ }} {\bf 108}, 011103 (2012)
  [arXiv:1109.0521 [astro-ph.HE]].
 
\bibitem{Aguilar:2013qda} 
  M.~Aguilar {\it et al.} [AMS Collaboration],
  {\it{Phys.\ Rev.\ Lett.\ }}  {\bf 110}, 141102 (2013).
  
\bibitem{Abeysekara:2017hyn} 
  A.~U.~Abeysekara {\it et al.},
  {\it{Astrophys.\ J.\ }}  {\bf 843}, no. 1, 40 (2017)
  [arXiv:1702.02992 [astro-ph.HE]].
  
\bibitem{Hooper:2017gtd} 
  D.~Hooper, I.~Cholis, T.~Linden and K.~Fang,
  {\it{Phys.\ Rev.\ }} D {\bf 96}, no. 10, 103013 (2017)
  [arXiv:1702.08436 [astro-ph.HE]].
  
\bibitem{Goodenough:2009gk} 
  L.~Goodenough and D.~Hooper,
  {\it{``Possible Evidence For Dark Matter Annihilation In The Inner Milky Way From The Fermi Gamma Ray Space Telescope,''}}
  arXiv:0910.2998 [hep-ph].
  
\bibitem{Hooper:2010mq} 
  D.~Hooper and L.~Goodenough,
  {\it{``Dark Matter Annihilation in The Galactic Center As Seen by the Fermi Gamma Ray Space Telescope,''}}
  {\it{Phys.\ Lett.\ }} B {\bf 697}, 412 (2011)
  [arXiv:1010.2752 [hep-ph]].
  
\bibitem{Hooper:2011ti} 
  D.~Hooper and T.~Linden
  {\it{``On The Origin Of The Gamma Rays From The Galactic Center,''}}
  {\it{Phys.\ Rev.\ }} D {\bf 84}, 123005 (2011)
  [arXiv: 1110.0006 [astro-ph.HE]].

\bibitem{Abazajian:2010zy} 
  K.~N.~Abazajian,
  {\it{``The Consistency of Fermi-LAT Observations of the Galactic Center with a Millisecond Pulsar Population in the Central Stellar Cluster,''}}
  {\it{JCAP}} {\bf 1103}, 010 (2011)
  [arXiv:1011.4275 [astro-ph.HE]].
  
\bibitem{Bartels:2015aea} 
  R.~Bartels, S.~Krishnamurthy and C.~Weniger,
  {\it{``Strong support for the millisecond pulsar origin of the Galactic center GeV excess,''}}
  {\it{Phys.\ Rev.\ Lett.\ }}  {\bf 116}, no. 5, 051102 (2016)
  [arXiv:1506.05104 [astro-ph.HE]].
  
\bibitem{Lee:2015fea} 
  S.~K.~Lee, M.~Lisanti, B.~R.~Safdi, T.~R.~Slatyer and W.~Xue,
  {\it{``Evidence for Unresolved $\gamma$-Ray Point Sources in the Inner Galaxy,''}}
  {\it{Phys.\ Rev.\ Lett.\ }}  {\bf 116}, no. 5, 051103 (2016)
  [arXiv:1506.05124 [astro-ph.HE]].
  
\bibitem{Bartels:2015aea} 
  R.~Bartels, S.~Krishnamurthy and C.~Weniger,
  {\it{Phys.\ Rev.\ Lett.\ }}  {\bf 116}, no. 5, 051102 (2016)
  [arXiv:1506.05104 [astro-ph.HE]].
  
\bibitem{Macias:2019omb} 
  O.~Macias, S.~Horiuchi, M.~Kaplinghat, C.~Gordon, R.~M.~Crocker and D.~M.~Nataf,
  {\it{JCAP}} {\bf 1909}, no. 09, 042 (2019)
  [arXiv:1901.03822 [astro-ph.HE]].
  
\bibitem{Leane:2019xiy} 
  R.~K.~Leane and T.~R.~Slatyer,
  arXiv:1904.08430 [astro-ph.HE].
  
\bibitem{Kovetz:2018zan} 
  E.~D.~Kovetz, V.~Poulin, V.~Gluscevic, K.~K.~Boddy, R.~Barkana and M.~Kamionkowski,
  {\it{Phys.\ Rev.\ }} D {\bf 98}, no. 10, 103529 (2018)
  [arXiv:1807.11482 [astro-ph.CO]].
  
\bibitem{Bernabei:2018yyw} 
  R.~Bernabei {\it et al.},
  {\it{Universe}} {\bf 4}, no. 11, 116 (2018)
  [{\it{Nucl.\ Phys.\ Atom.\ Energy }} {\bf 19}, no. 4, 307 (2018)]
  [arXiv:1805.10486 [hep-ex]].
  
\bibitem{Alexander:2016aln} 
  J.~Alexander {\it et al.},
  arXiv:1608.08632 [hep-ph].

\bibitem{Battaglieri:2017aum} 
  M.~Battaglieri {\it et al.},
  arXiv:1707.04591 [hep-ph].
  
\bibitem{Abramoff:2019dfb} 
  O.~Abramoff {\it et al.} [SENSEI Collaboration],
  {\it{Phys.\ Rev.\ Lett.\ }}  {\bf 122}, no. 16, 161801 (2019)
  [arXiv:1901.10478 [hep-ex]].
  
\bibitem{Salemi:2019xgl} 
  C.~P.~Salemi [ABRACADABRA Collaboration],
  arXiv:1905.06882 [hep-ex].
  
\bibitem{Donini:2018tsg} 
  A.~Donini, S.~Palomares-Ruiz and J.~Salvado,
  {\it{Nature Phys.\ }}{\bf 15}, no. 1, 37 (2019)
  [arXiv:1803.05901 [hep-ph]].
  

 \end{thebibliography}
\end{document}